\documentclass[11pt,a4paper]{article}
\usepackage[hyperref]{acl2020}
\usepackage{times}
\usepackage{latexsym,enumitem}

\usepackage{graphicx}
\usepackage{subfig}
\usepackage{xspace}
\usepackage{multirow} 
\usepackage{xcolor,colortbl}   

\newcommand{\CodeFont}[1]{\texttt{\small #1}}
\newcommand{\cc}{\CodeFont{CrossCheck}\@\xspace}
\newcommand{\ignore}[1]{}
\newcommand{\nop}[1]{}

\newcommand{\etc}{\textit{etc.~}}
\newcommand{\eg}{\textit{e.g.,~}}
\newcommand{\ie}{\textit{i.e.,~}} 
\newcommand{\fs}{F1-score\@\xspace}

\usepackage{microtype}

\aclfinalcopy 

\title{CrossCheck: Rapid, Reproducible, and Interpretable Model Evaluation}

\author{
  Dustin Arendt$^*$  ~~~~~ Zhuanyi Huang$^*$    \\
  \bf ~~~~~ 
  Prasha Shrestha\dag   ~~~~~ Ellyn Ayton\dag   ~~~~~ Maria Glenski\dag  ~~~~~  Svitlana Volkova\dag  \\ 
  $^*$Visual Analytics Group, \dag Data Sciences and Analytics Group \\ 
  Pacific Northwest National Laboratory \\
  \texttt{\{first\}.\{last\}@pnnl.gov} 
  \\ 
} 

\date{}

\begin{document}
\maketitle
\begin{abstract}
Evaluation beyond aggregate performance metrics, e.g. \fs, is crucial to both establish an appropriate level of trust in machine learning models and identify future model improvements.
In this paper we demonstrate {\cc}, an interactive visualization tool for rapid cross-model comparison and reproducible error analysis.
We describe the tool and discuss design and implementation details. We then present three use cases  (named entity recognition, reading comprehension, and clickbait detection) that show the benefits of using the tool for model evaluation. \cc allows data scientists to make informed decisions to choose between multiple models, identify when the models are correct and for which examples, investigate whether the models are making the same mistakes as humans, evaluate models' generalizability and highlight models' limitations, strengths and weaknesses.
Furthermore, \cc is implemented as a Jupyter widget, which allows rapid and convenient integration into data scientists' model development workflows.
\end{abstract}

\section{Motivation}
    Complex machine learning (ML) models for NLP are imperfect, opaque, and often brittle. Gaining an effective understanding and actionable insights about model strengths and weaknesses is challenging because simple metrics like accuracy or \fs are not sufficient to capture the complex relationships between model inputs and outputs. Therefore, standard performance metrics should be augmented with exploratory model performance analysis, where a user can interact with inputs and outputs to find patterns or biases in the way the model makes mistakes to answer the questions of when, how, and why the model fails. 
Many researchers agree that ML models have to be optimized not only for expected task performance but for other important criteria such as explainability, interpretability, reliability, and fairness that are prerequisites for trust \citep{lipton2016mythos,doshi2017towards,poursabzi2018manipulating}.

To support ML model evaluation beyond standard performance metrics we developed a novel interactive tool \cc\footnote{\url{https://github.com/pnnl/crosscheck}}. Unlike several recently developed tools for analyzing model errors~\citep{agarwal2014error,wu2019errudite}, understanding model outputs~\citep{lee2019qadiver,hohman2019gamut} and model interpretation and diagnostics~\citep{kahng2017actiVis,kahng2016visual,zhang2018manifold}, \cc is designed to allow rapid prototyping and cross-model comparison to support comprehensive experimental setup and gain interpretable and informative insights into model performance.  

Many visualization tools have been developed recently, \eg  ConvNetJS\footnote{\url{https://github.com/karpathy/convnetjs}}, TensorFlow Playground\footnote{\url{https://playground.tensorflow.org/}}, that focus on structural interpretability~\citep{kulesza2013too,hoffman2018explaining} and operate in the neuron activation space to explain models' internal decision making processes~\citep{kahng2017actiVis} or focus on visualizing a model's decision boundary to increase user trust~\citep{ribeiro2016should}. Instead, \cc targets functional interpretability and operates in the model output space to diagnose and contrast model performance.

\begin{figure*}[h!]

    \centering
    \includegraphics[width=0.9\textwidth]{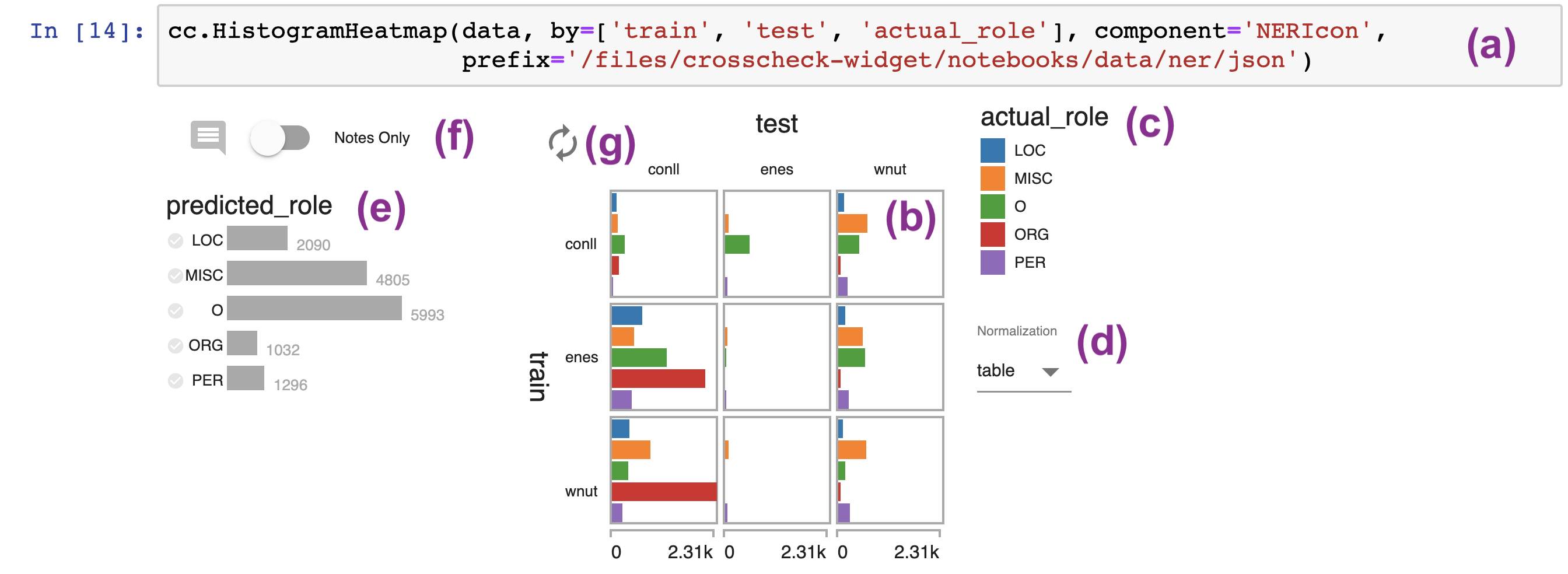}
    \caption{\cc embedded in a Jupyter Notebook cell: (a) code used to instantiate the widget (b) the histogram heatmap shows the distribution of the third variable for each combination of the first two (c) the legend for the third variable (d) normalization controls (e) histogram \& filter for remaining variables (f) controls for notes (g) button to transpose the rows and columns.}
    \label{fig:crosscheck}
\end{figure*}

Similar work to \cc includes AllenNLP Interpret~\citep{wallace2019allennlp} and Errudite~\cite{wu2019errudite}. AllenNLP Interpret
relies on saliency map visualizations to uncover model biases, find decision rules, and diagnose model errors. Errudite
implements a domain specific language for counterfactual explanations. Errudite and AllenNLP Interpret focus primarily on error analysis for a single model, while our tool is specifically designed for contrastive evaluation across multiple models \eg neural architectures with different parameters, datasets, languages, domains, \etc

Manifold~\citep{zhang2018manifold} supports cross-model evaluation, however the tool is narrowly focused on model confidence and errors via pairwise model comparison with scatter plots. \cc enables users to investigate ``where'' and ``what'' types of errors the model makes and, most importantly, assists the user with answering the question ``why" the model makes that error by relying on a set of derived attributes from the input like inter-annotator agreement, question type, length of the answer, the input paragraph, \etc

Before implementing \cc our error analysis process was manual, time-consuming, \textit{ad hoc}, and difficult to reproduce. Thus, we endeavored to build a tool to make our process faster and more principled, but based on the successful error analysis techniques we had practiced. \cc helps to  calibrate users' trust by enabling users to:
\begin{itemize}[noitemsep,nolistsep]
    \item choose between multiple models,
    \item see when the model is right (or wrong) and further examine those examples,
    \item investigate whether the model makes the same mistakes as humans,
    \item highlight model limitations, and
    \item understand how models generalize across domains, languages and datasets.
\end{itemize}

\section{CrossCheck}

\cc's input is a single mixed-type table, {\it i.e.} a pandas DataFrame\footnote{\url{http://pandas.pydata.org}}. It is embedded in a Jupyter\footnote{\url{https://jupyter.org}} notebook to allow for tight integration with data scientists' workflows (see Figure~\ref{fig:crosscheck}a). 
Below we outline the features of \cc in detail.

\cc's main view (see Figure~\ref{fig:crosscheck}b) extends the \textit{confusion matrix} visualization technique by replacing each cell in the matrix with a histogram~---~we call this view the histogram heatmap.
Each cell shows the distribution of a third variable conditioned on the values of the corresponding row and column variables. Every bar represents a subset of instances, \ie rows in the input table, and encodes the relative size of that group.
This view also contains a legend showing the bins or categories for this third variable (see Figure~\ref{fig:crosscheck}c).

\begin{figure*}[t!]
    \centering
    \vspace{-0.7cm}
    \subfloat[by table]{\includegraphics[height=5cm]{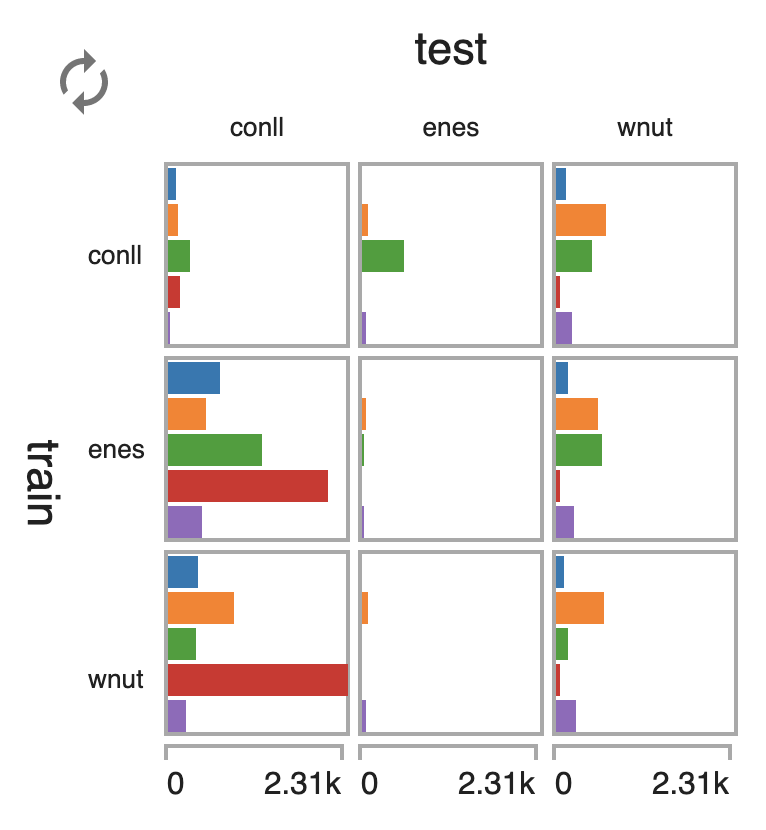}}
    \subfloat[by column]{\includegraphics[height=5cm]{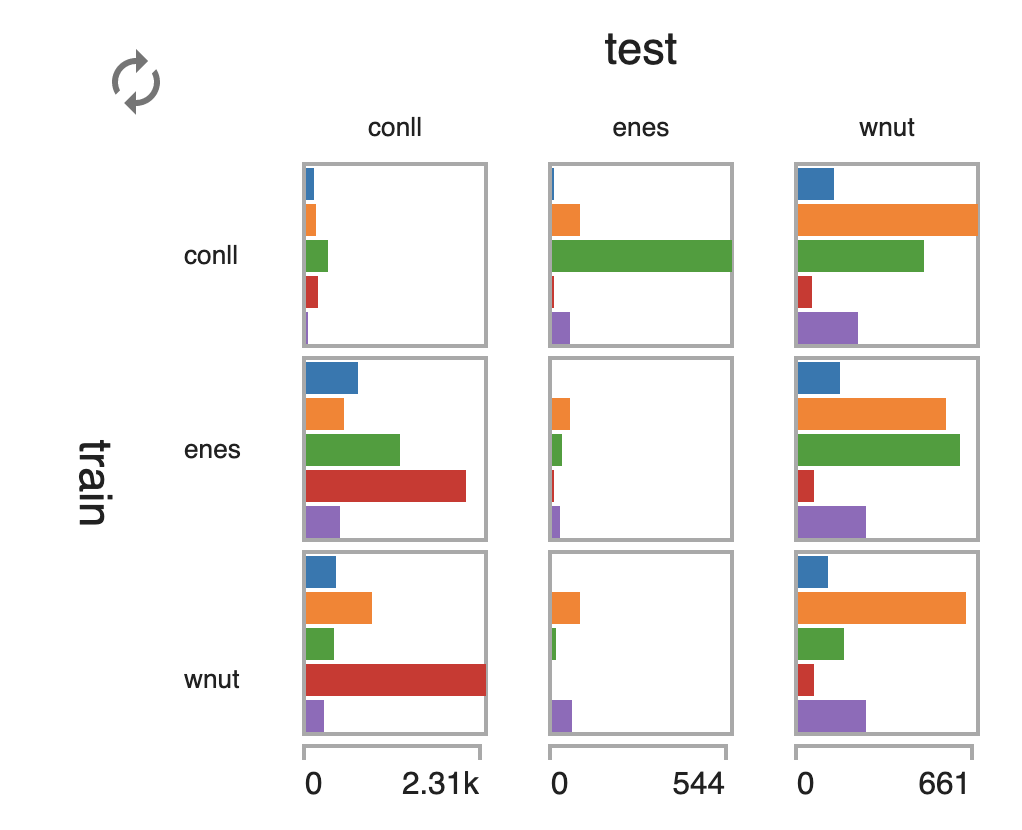}}
    \subfloat[by cell]{\includegraphics[height=5cm]{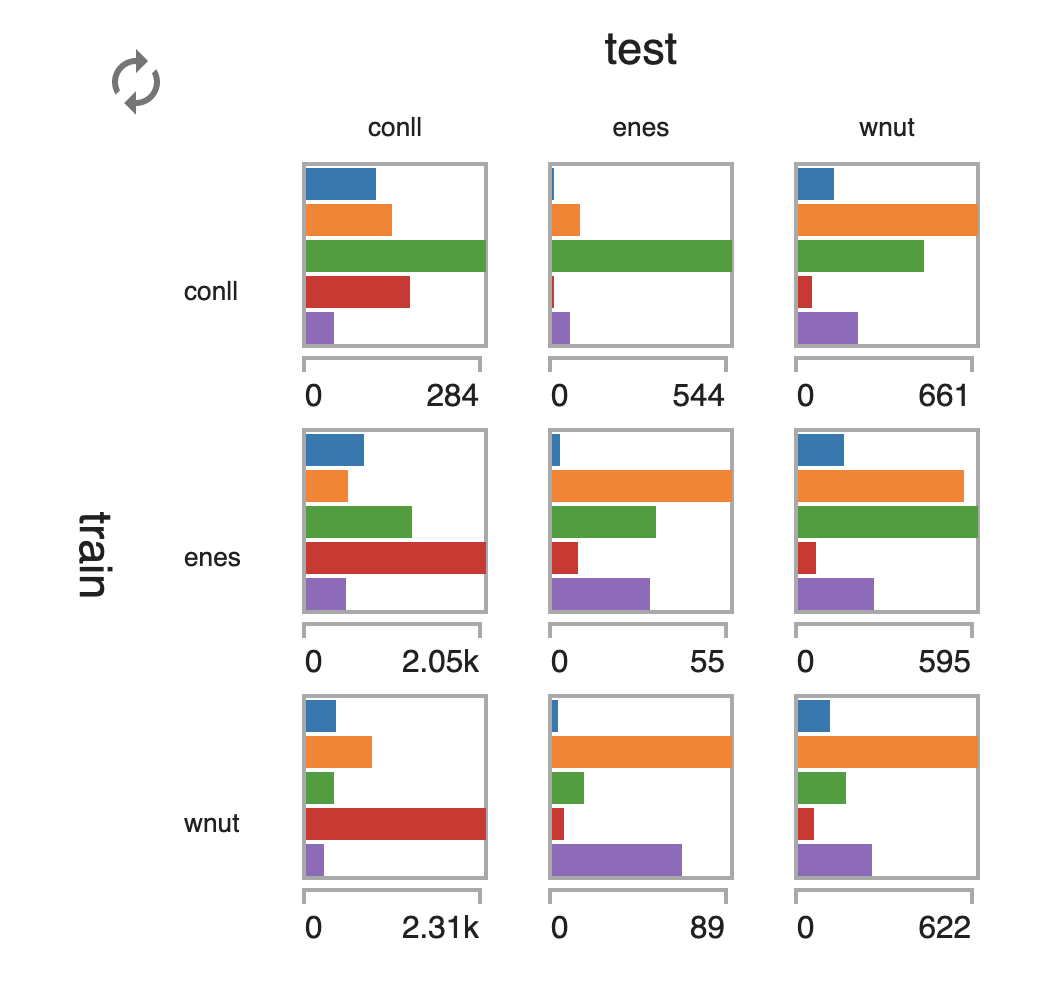}}
    \caption{\cc supports three histogram normalization options that affect how axes and padding are rendered to improve the readability and interpretation of the view (a) by table: minimal padding, the same x-axes are shown on the bottom row  (b) by column: extra padding between columns, different x-axes are shown on the bottom row (c) by cell: extra padding between rows and columns, different x-axes are shown for each cell.}
    \label{fig:cc-normalization}
    \vspace{-0.2cm}
\end{figure*}

The histograms in each cell in \cc are drawn horizontally to encourage comparison across cells in the vertical direction.
\cc supports three normalization schemes (see Figure~\ref{fig:crosscheck}d), \ie setting the maximum x-value in each cell: normalizing by the maximum count within the entire matrix, within each column, or within each cell.
To emphasize the current normalization scheme, we also selectively show certain axes and adjust the padding between cells.
Figure~\ref{fig:cc-normalization} illustrates how these different normalization options appear in \cc.
By design, there is no equivalent row normalization option, but the matrix can be transposed (see Figure~\ref{fig:crosscheck}g) to swap the rows and columns for an equivalent effect.

Any variables not directly compared in the histogram heatmap are visualized on the left side of the widget as histograms (see Figure~\ref{fig:crosscheck}e). These histograms also allow the user to filter data when it is rendered in the main view by clicking on the bar(s) corresponding to the data they want to keep.

We also allow users to take notes on instances to support their evaluation workflow. Clicking the switch labeled ``Notes Only'' (see Figure~\ref{fig:crosscheck}f) filters out all instances that are not annotated in the main view, showing the user what has been annotated in the context of the current variable groupings.

\section{Use Cases and Evaluation}
    In this section, we highlight how \cc can be used in core NLP tasks such as named entity recognition (NER) and reading comprehension (RC) or practical applications of NLP such as clickbait detection (CB). We present an overview of the datasets used for each task below:
\begin{itemize}[noitemsep,nolistsep]
    \item NER: CoNLL~\cite{sangmeulder2003conll}, ENES~\cite{aguilar2018named}, WNUT 17 Emerging Entities~\cite{derczynski2017results}\footnote{\url{github.com/leondz/emerging_entities_17}},
    \item MC: Stanford Question Answering Dataset (SQuAD)~\cite{rajpurkar2016squad}\footnote{ \url{rajpurkar.github.io/SQuAD-explorer/}}, 
    \item CB: Clickbait Challenge 2017~\cite{potthast2018crowdsourcing}\footnote{ \url{www.clickbait-challenge.org/\#data}}.
\end{itemize}

\ignore{
\begin{table}[th!]
\caption{Summary of publicly available datasets used for each use case.}
\centering
\small
\begin{tabular}{p{0.5cm} p{5.5cm} l} 
	\hline
	& Dataset & Link\\
	\hline 
	NER &  CoNLL~\cite{sangmeulder2003conll}  \\
	& ENES~\cite{aguilar2018named} & \\
	&  WNUT 17 Emerging Entities~\cite{derczynski2017results} & 
	
	\rowcolor{lightgray!20}
	\multirow{3}{*}{MC} &  Stanford Question Answering Dataset~\cite{rajpurkar2016squad} & \url{rajpurkar.github.io/SQuAD-explorer/} \\
	
	\rowcolor{lightgray!20}
	& TriviaQA~\cite{joshiTriviaQA2017} & \url{nlp.cs.washington.edu/triviaqa/}
	 \\
	 
	\multirow{2}{*}{FC} &  LIAR Benchmark Dataset for Fake News Detection~\cite{wang2017liar} & \url{github.com/thiagorainmaker77/liar_dataset} \\
	\multirow{2}{*}{CB}
	&  Clickbait Challenge 2017~\cite{potthast2018crowdsourcing} & \url{www.clickbait-challenge.org/#data} \\
	\hline
\end{tabular}
\label{tab:datasets}
\end{table}
}

\subsection{Named Entity Recognition (NER)}
\begin{figure}[t!]
    \centering
    \subfloat[\small Named Entity Recognition]{\includegraphics[width=0.45\textwidth]{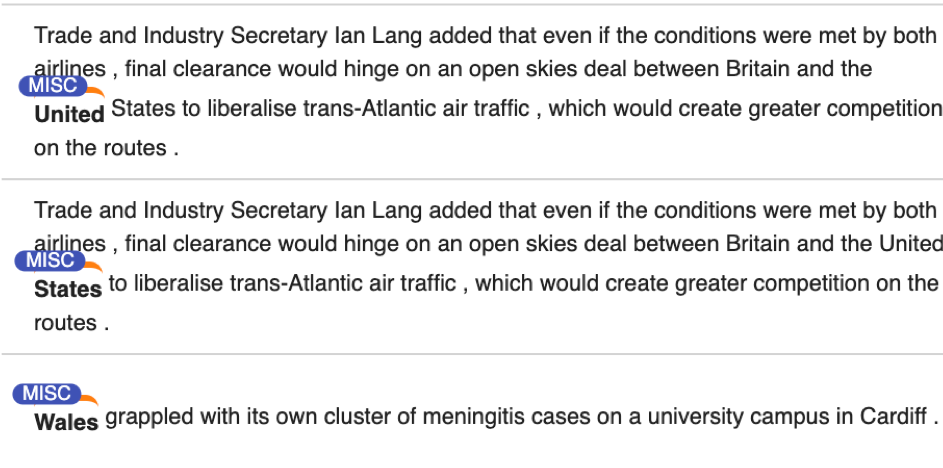}}
    
    \subfloat[\small Reading Comprehension]{\includegraphics[width=0.5\textwidth]{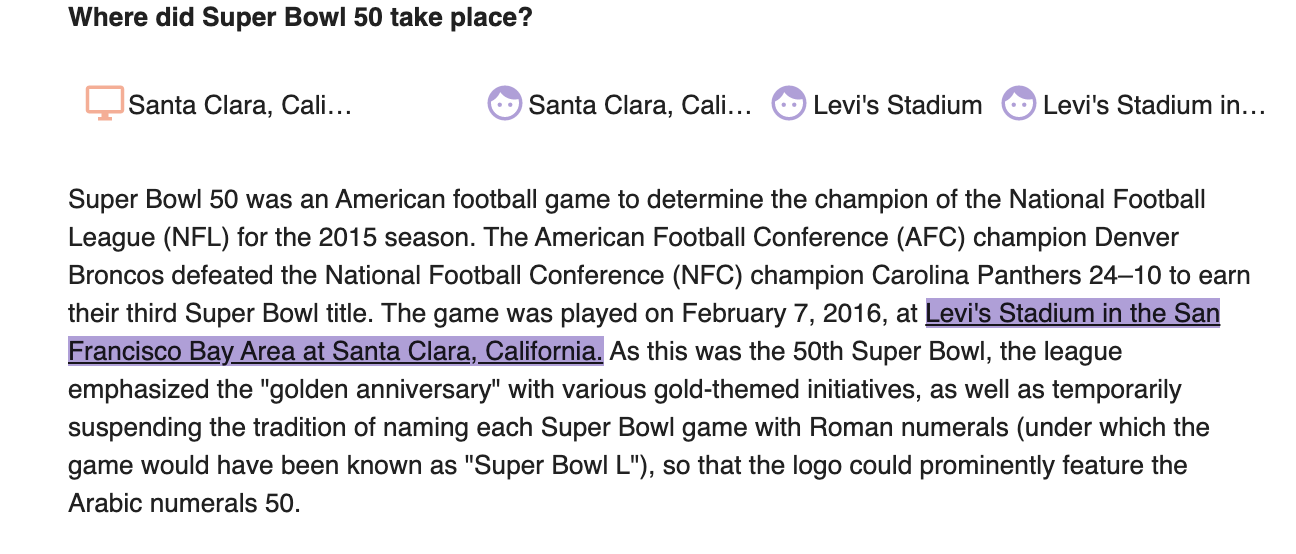}}
     
    \caption{Examples of model outputs in \cc for core NLP tasks -- for the NER task (above), predicted named entities are highlighted, and for the RC task (below), predicted answer span is highlighted.} 
    \label{fig:example_outputs}
    \vspace{-0.3cm}
\end{figure}

To showcase \cc, we trained and evaluated the AllenNLP NER model~\cite{peters2017semi} across three benchmark datasets -- CoNLL, WNUT, and ENES, producing nine different evaluations. 
The model output includes, on a per-token level, the model prediction, the ground truth, the original sentence (for context), and what the training and testing datasets were as shown in Figure~\ref{fig:example_outputs}a.

\begin{figure*}[t!]
    \centering
    \includegraphics[width=\textwidth]{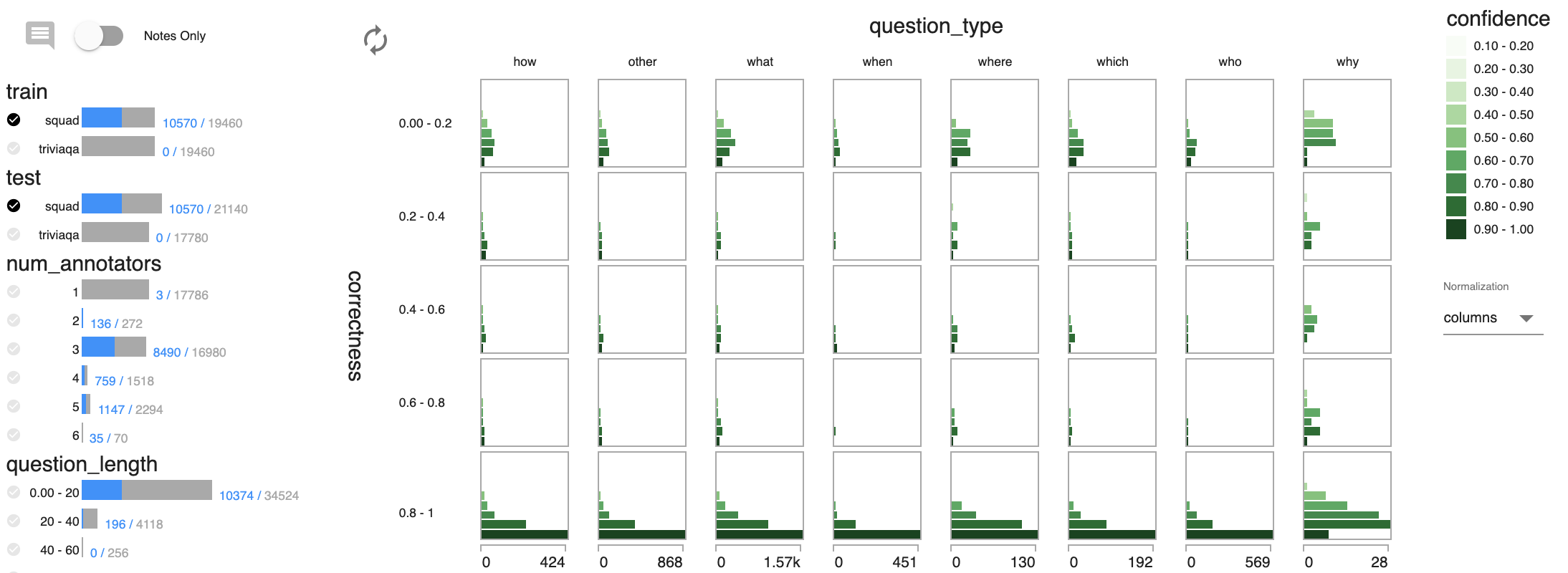}
    \caption{\cc for evaluation of reading comprehension models to understand the relationship between correctness, confidence and question types. This highlight models limitations and shows for what examples the model answers correctly.}
    \label{fig:mc}
\end{figure*}

This experiment was designed to let us understand how models trained on different datasets generalize to the same test data (shown in columns), how models trained on the same training data transfer to predict across different test datasets (shown in rows). Figure~\ref{fig:cc-normalization} illustrates the \cc grid of train versus test datasets. The data has been filtered so that only errors contribute to the bars so we can see a distribution of errors per train-test combination across the actual role. Since the CoNLL dataset is much larger, we can allow normalization within columns in Figure~\ref{fig:cc-normalization}b to look for patterns within those sub-groups.  

For the same experimental setup, Table~\ref{lab:NER_F1} summarizes performance with  F1-scores. Unlike the F1-score table, \cc reveals that models trained on social media data misclassify ORG on the news data, and the news models overpredict named entities on social media data.

\begin{table}
\small
\centering
\caption{Traditional evaluation: F1-scores for the NER models trained and tested across domains.}
    \begin{tabular}{l c c c}
        \hline
        Train $\backslash$ Test 
        & CoNLL  & WNUT  & ENES  \\
        \hline
        CoNLL & 92.51 & 40.10 & 11.88 \\
        WNUT & 55.75 & 44.73 & 33.33\\
        ENES & 50.78 & 57.48 & 64.00 \\
        \hline
    \end{tabular}
    \label{lab:NER_F1}
\end{table}

\subsection{Reading Comprehension (RC)}
Similar to NER, we trained an AllenNLP model for reading comprehension~\cite{seo2016bidirectional} that is designed to find the most relevant span for a question and paragraph input pair. The model output includes, on a question-paragraph level: the model prediction span, ground truth span, model confidence, question type and length, the number of annotators per question, and what the train and test datasets were, as shown in Figure~\ref{fig:example_outputs}b.\footnote{We evaluated RC on SQuAD and TriviaQA datasets, but with space limitations only present results for SQuAD.} Figure~\ref{fig:mc} presents the \cc view of the model's correctness and confidence across question types. We can see that across all types of questions when the model is correct it has higher confidence (bottom row), and lower confidence when incorrect (top row). In addition, we see model behavior has higher variability when predicting ``why'' questions compared to other types.

\ignore{ 
\subsection{Fact-checking}

- identifying biases

\begin{table}
\small
\caption{Traditional evaluation: F1-scores per class and the macro average and weighted F1-scores overall for each models predictions compared to ground truth labels.}
    \begin{tabular}{l c c}
        \hline
        {} & & Biased  \\
        {} &  Weighted Random &  Weighted Random \\
        \hline
        true         &                0.21 &             0.09 \\
        mostly-true  &                0.22 &             0.15 \\
        half-true    &                0.24 &             0.30 \\
        barely-true  &                0.17 &             0.14 \\
        false        &                0.20 &             0.13 \\
        pants-fire   &                0.07 &             0.06 \\
        \hline
        macro avg    &                0.18 &             0.15 \\
        weighted avg &                0.20 &             0.16 \\
        \hline
    \end{tabular}
\end{table}

\begin{figure}[t]
    \centering  
    
    \includegraphics[width=0.48\textwidth]{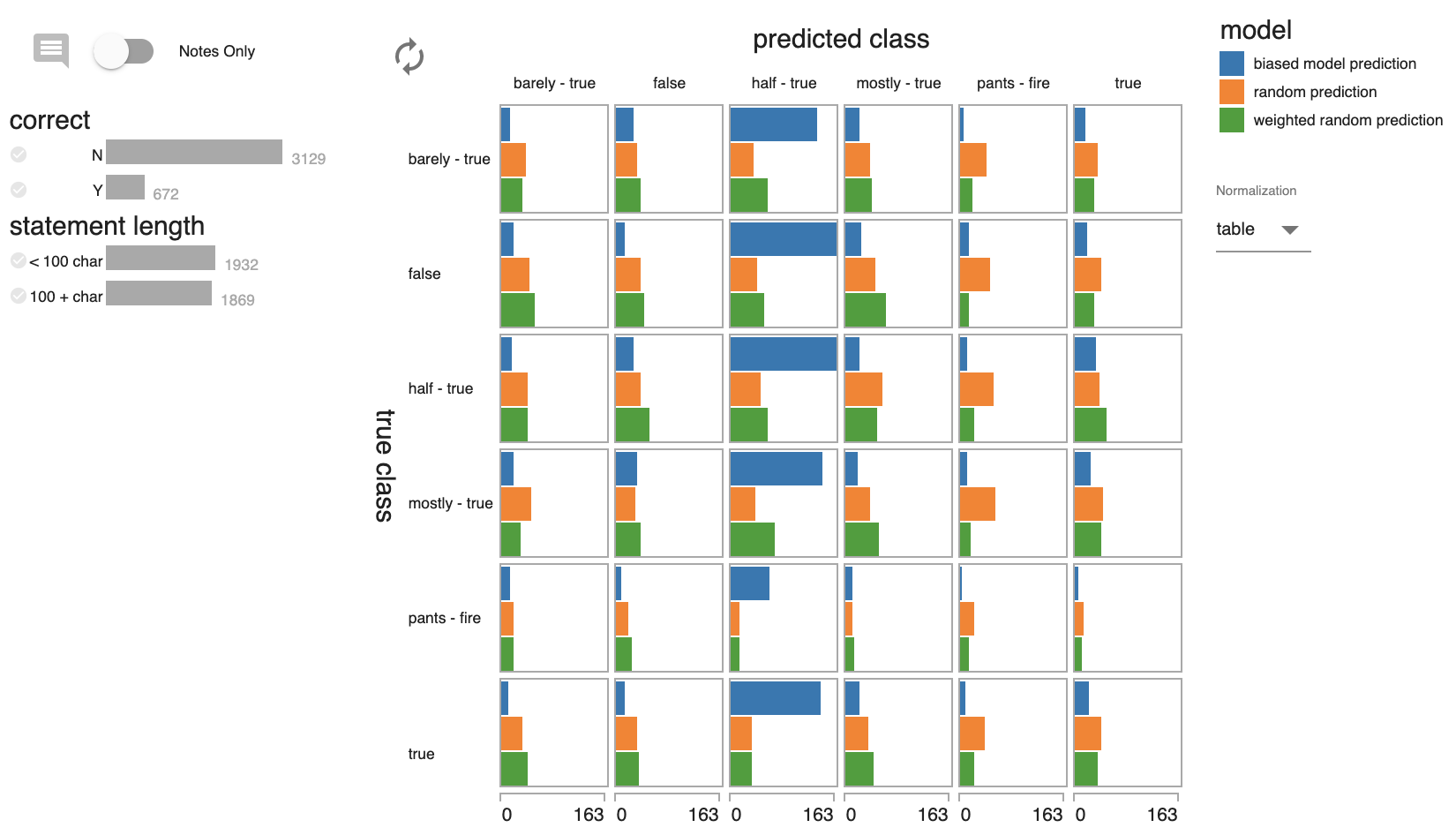}\\
    \hspace{\baselineskip}
    \includegraphics[width=0.48\textwidth]{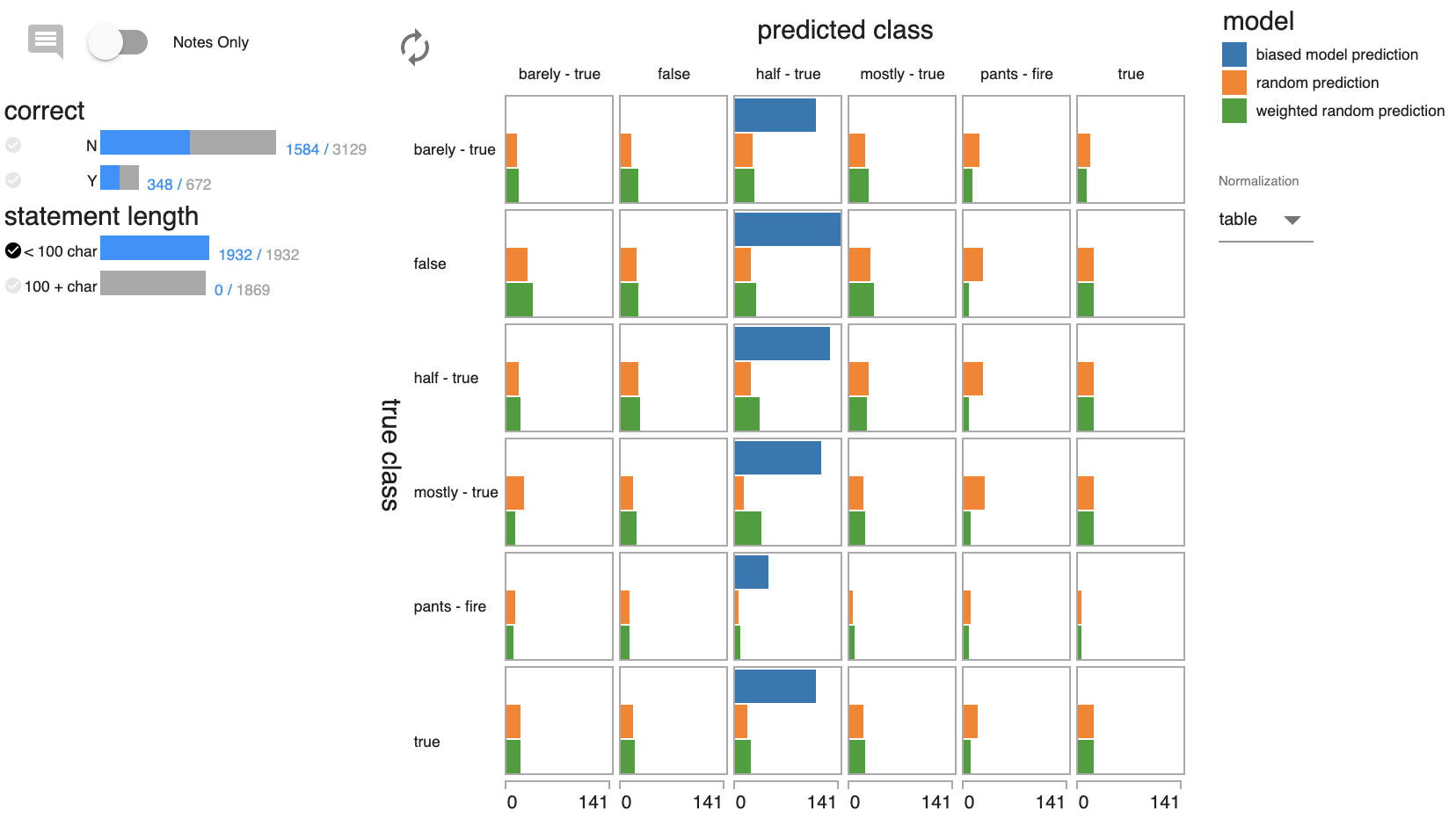}
         
    \caption{Model comparison across LIAR models, highlighting bias in the biased weighted random predictions with histograms normalized across the full table.}
    \label{fig:cb_team_filter}
\end{figure}
}

\subsection{Clickbait Detection}
\label{sec:clickbait} 
Finally, we demonstrate \cc for comparison of regression models. We use a relevant application of NLP in the domain of deception detection (clickbait detection) that was the focus of the Clickbait Challenge 2017, a shared task focused on identifying a score (from 0 to 1) of how ``clickbait-y'' a social media post (\ie tweet on Twitter) is, given the content of the post (text and images) and the linked article webpages. We use the validation dataset that contains 19,538 posts (4,761 identified as clickbait) and pre-trained models released on GitHub after the challenge by two teams ({\it blobfish} and {\it striped-bass)}\footnote{Models and code available via \url{github.com/clickbait-challenge/} repositories.}. 
\begin{figure*}[t!]
    \centering 
    \includegraphics[width=0.9\textwidth]{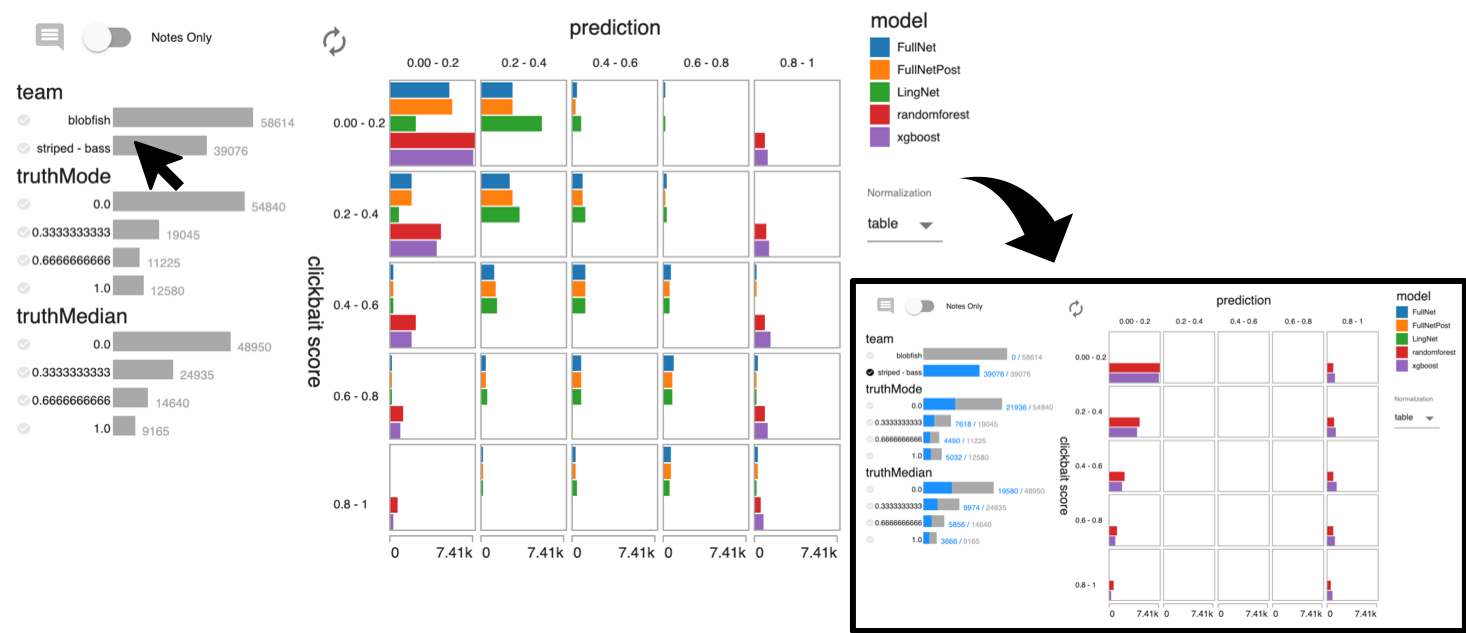}

    \caption{\cc for cross-model comparison across two teams who competed in the Clickbait Challenge 2017 shared task, highlighting distinctions in the variety of prediction outputs with histograms normalized across the full table that become particularly clear when team filters are selected.}
    \label{fig:cb_team_filter}
\end{figure*}

In Figure~\ref{fig:cb_team_filter} we illustrate how \cc can be used to compare across multiple models and across multiple classes of models.\footnote{Note, models could also be grouped by any number of shared characteristics such as the algorithms or architectures used (\eg different neural architectures used in deep learning models, or models that use deep learning versus those that do not), hyper-parameter settings, granularity of prediction outputs, ensembles versus single models, \etc} When filtered to show only the {\it striped-bass} models (shown at right), a strategy to predict coarse (0 or 1) clickbait scores versus fine-grained clickbait scores is clearly evident in the {\it striped-bass} model predictions. Here, there is a complete lack of predictions falling within the center three columns so even with no filters selected (shown at left), \cc provides indications that there may be this disparity in outcomes between models (an explanation for the disparity in F1-scores in Table~\ref{tab:f1clickbait}. In cases where there is a more nuanced or subtle disparity, shallow exploration with different filters within \cc can lead to efficient, effective identification of these key differences in model behavior.

\begin{table}[t]
\centering
\small
\caption{Traditional evaluation summary table contrasting mean squared error (MSE) and mean absolute error (MAE) of each model's predictions.} 
    \begin{tabular}{llrr}
        \hline
                 Team &         Model &    MSE &    MAE \\
        \hline
        \multirow{3}{*}{blobfish} &   FullNetPost &  0.026 &  0.126 \\
              &       FullNet &  0.027 &  0.130 \\
              &       LingNet &  0.038 &  0.157 \\
        \hline
         \multirow{2}{*}{striped-bass} &       xgboost &  0.171 &  0.326 \\
         &  randomforest &  0.180 &  0.336 \\
        \hline
    \end{tabular}
    \label{tab:f1clickbait}
    \vspace{-0.2cm}
\end{table}

\ignore{
\begin{figure*}
    \centering 
    \subfloat[No Filters]{\includegraphics[height=3in]{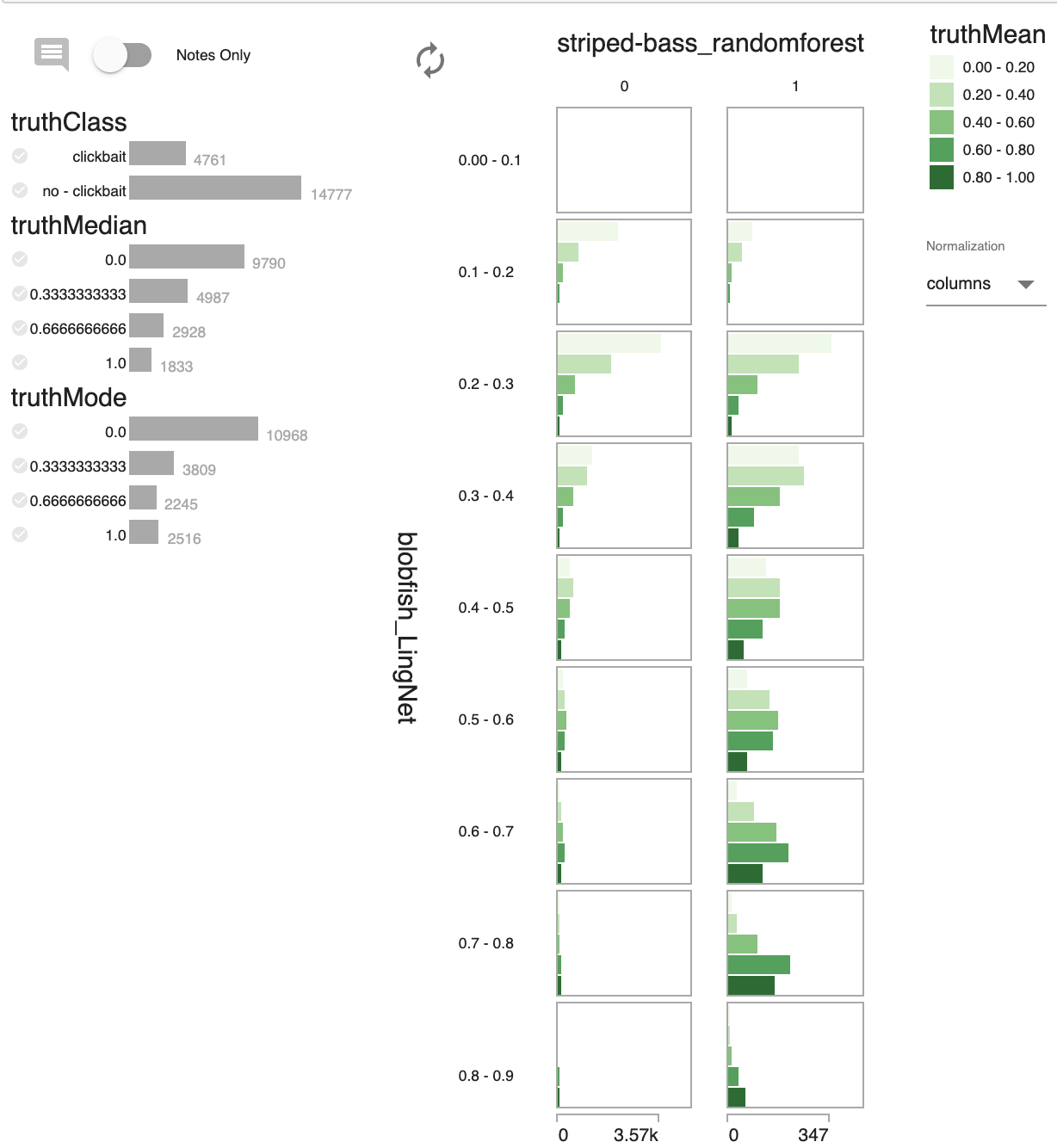}} 
    \hspace{2\baselineskip}
    \subfloat[Filter to "No Clickbait"]{\includegraphics[height=3in]{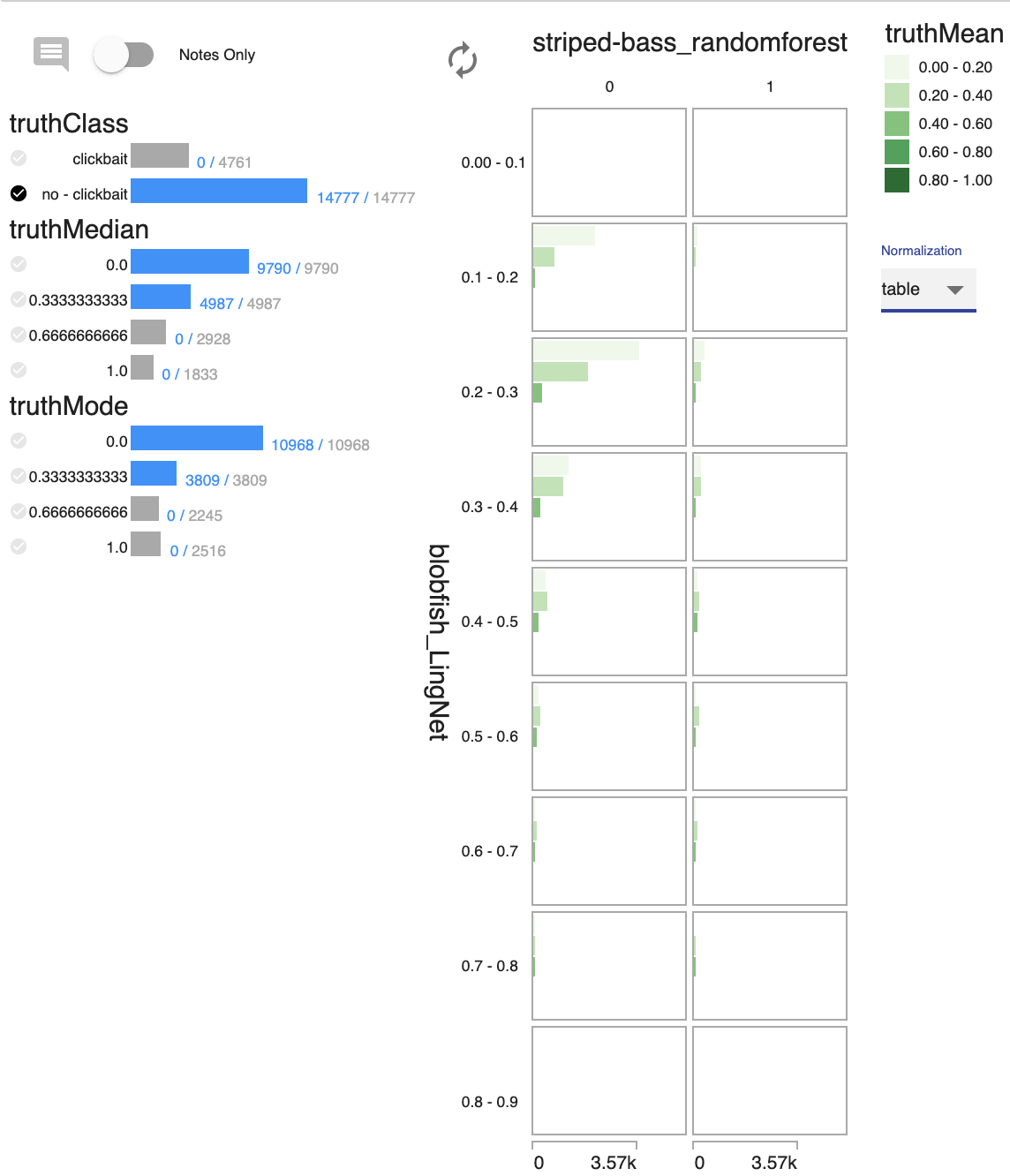}} 
    \caption{Model Comparison Supported by \cc}
    \label{fig:my_label}
\end{figure*}
}

\section{Design and Implementation}
    \label{sec:design}
We designed \cc following a user-centered design methodology.
This is a continuous, iterative process where we identify needs and goals, implement prototypes, and solicit feedback from our users to incorporate in the tool.
Our users were data scientists, specifically NLP researchers and practitioners, tasked with the aforementioned model evaluation challenge.
We identified \cc's goals as allowing the user to: understand how instance attributes relate to model errors; provide convenient access to raw instance data; integrate into a data scientists workflow; and reveal and understand disagreement across models, and support core NLP tasks and applications.

\newcommand{\DesignRound}[4]{
\paragraph{Round #1---#2}#3~\textit{User feedback: #4}
}

\subsection{Design Iterations}

\DesignRound{1}{Heatmaps (functional prototype)}{Our first iteration extended the confusion matrix visualization technique with a functional prototype that grouped the data by one variable, and showed a separate heatmap for each distinct value in that group.
}{though heatmaps are familiar, the grouping made the visualization misleading and difficult to learn.}

\DesignRound{2}{Table \& Heatmap (wireframes)}{We wireframed a standalone tool with histogram filters, a sortable table, and a more traditional heatmap visualization with a rectangular brush to reveal raw instance data.}{the sortable table and brushing would be useful, but the heatmap has essentially the same limitations as confusion matrices.}

\DesignRound{3}{Histogram Heatmap (wireframes)}{We wireframed a modified heatmap where each cell was replaced with a histogram showing the distribution of a third variable conditioned on the row and column variables. This modified heatmap was repeated for each variable in the dataset except for the row and column variables.}{Putting the histogram inside the heatmap seems useful, but multiple copies would be overwhelming and too small to read. We would prefer to work with just one histogram heatmap.}

\DesignRound{4}{\cc (functional prototype)}{We implemented a single ``histogram heatmap'' inside a Jupyter widget, and made raw instance data available to explore by clicking on any bar. Additionally we incorporated histogram filters from the Round 2 design and allowed the user to change the histogram normalization.}{the tool was very useful, but could use minor improvements e.g., labeled axes and filtering, as well as ability to add annotation on raw data.}

\DesignRound{5}{\cc (polished prototype)}{We added minor features like a legend, a matrix transpose button, axis labels, dynamic padding between rows and columns (based on normalization), and the ability to annotate instances with notes.}{the tool works very well, but screenshots aren't suitable to use in publications.}

    \subsection{Implementation Challenges}
\label{sec:jupyter-bottleneck}
To overcome the rate limit between the python kernel and the web browser (see the \CodeFont{NotebookApp.iopub\_data\_rate\_limit} Jupyter argument) our implementation separates raw instance data from tabular data to be visualized in \cc's histogram heatmap. The tool groups tabular data by each field in the table and passed as a list of each unique field/value combinations and the corresponding instances within that bin. This is computed efficiently within the python kernel (via a pandas \CodeFont{groupby}). This pre-grouping reduces the size of the payload passed from the python kernel to the web browser and allows for the widget to behave more responsively because visualization and filtering routines do not need to iterate over every instance in the dataset.
The tool stores raw instance data as individual JSON files on disk in a path visible to the Jupyter notebook environment. When the user clicks to reveal raw instance data, this data is retrieved asynchronously using the web browser's XMLHttpRequest (XHR). This allows the web browser to only retrieve and render the few detailed instances the user is viewing at a time.
\section{Discussion}

\cc is designed to quickly and easily explore a myriad of combinations of characteristics of both models (\eg parameter settings, network architectures) and datasets used for training or evaluation. It also provides users the ability to efficiently compare and explore model behavior in specific situations and \textit{generalizability of models} across datasets or domains. Most importantly, \cc can easily generalize to evaluate models on unlabeled data based on model agreement.

With its simple and convenient integration into data scientists' workflows, \cc enables users to perform extensive error analysis in an efficient and reproducible manner. The tool can be used to evaluate across models trained on image, video, tabular data, or combinations of data types with interactive exploration of specific instances (\eg those responsible for different types of model errors) on demand.

\paragraph{Limitations}
While pairwise model comparison with \cc is straightforward by assigning each model to a row and column in the histogram heatmap, comparing more than two models requires concessions.
Effective approaches we have taken for $n$-way comparisons include computing an agreement score across the models per instance or using a long instead of wide table format (as was used in Figure~\ref{fig:cb_team_filter}) that is less efficient.

Our users also had difficulty directly incorporating findings in \cc into scientific publications, due to a tradeoff between effective exploration versus communication.
In this case the major concerns were that text and axes that were designed for quick, at-a-glance consumption were not appropriate after screen capturing and insertion into documents for publication.

\paragraph{Future Work}
Another challenge with the tool is that adding visualizations for new use cases requires custom JavaScript code to be written, requiring end-users to work with a development version of the tool. Future work may include writing a generic set of components that cover the basics for most potential NLP use cases, or otherwise allow the tool to be extended with custom JavaScript source without re-compiling the packages. 

\section{Conclusions}

We have presented \cc\footnote{\url{https://github.com/pnnl/crosscheck}}, a new interactive visualization tool that enables rapid, interpretable model evaluation and error analysis. There are several key benefits to performing evaluation and analyses using our tool, especially compared to {\ie adhoc} or manual approaches because \cc:
\begin{itemize}[noitemsep]
    \item is generalizable across text, images, video, tabular, or combinations of multiple data types,
    \item can be integrated directly into existing workflows for rapid and highly reproducible error analysis during and after model development,
    \item users can interactively explore errors conditioning on different model/data features, and
    \item users can view specific instances of inputs that cause model errors or other interesting behavior within the tool itself.
\end{itemize}

\section*{Acknowledgments}
The project depicted was sponsored by the Department of the Defense, Defense Threat Reduction Agency. The content of the information does not necessarily reflect the position or the policy of the Federal Government, and no official endorsement should be inferred.
Subsequent evaluation of the tool described in this paper was conducted under the Laboratory Directed Research and Development Program at Pacific Northwest National Laboratory, a multiprogram national laboratory operated by Battelle for the U.S. Department of Energy. 


\bibliography{refs}
\bibliographystyle{acl_natbib}

\newpage

\appendix

\section{Appendices} \subsection{Preliminary Designs}
\begin{figure*}[h]
\subfloat[Table]{\includegraphics[width=.495\textwidth]{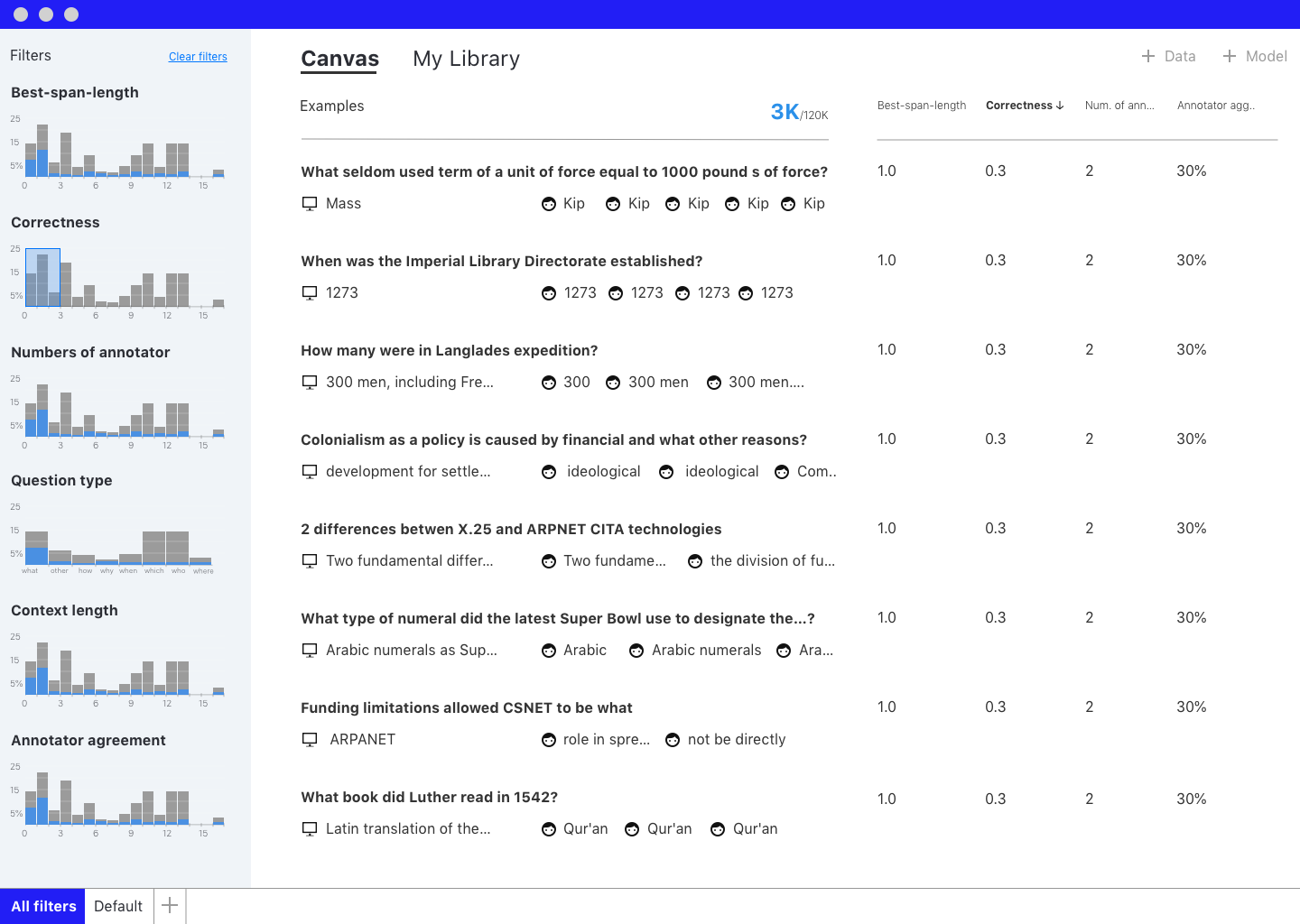}}\hspace{.01\textwidth}
\subfloat[Heatmap]{\includegraphics[width=.495\textwidth]{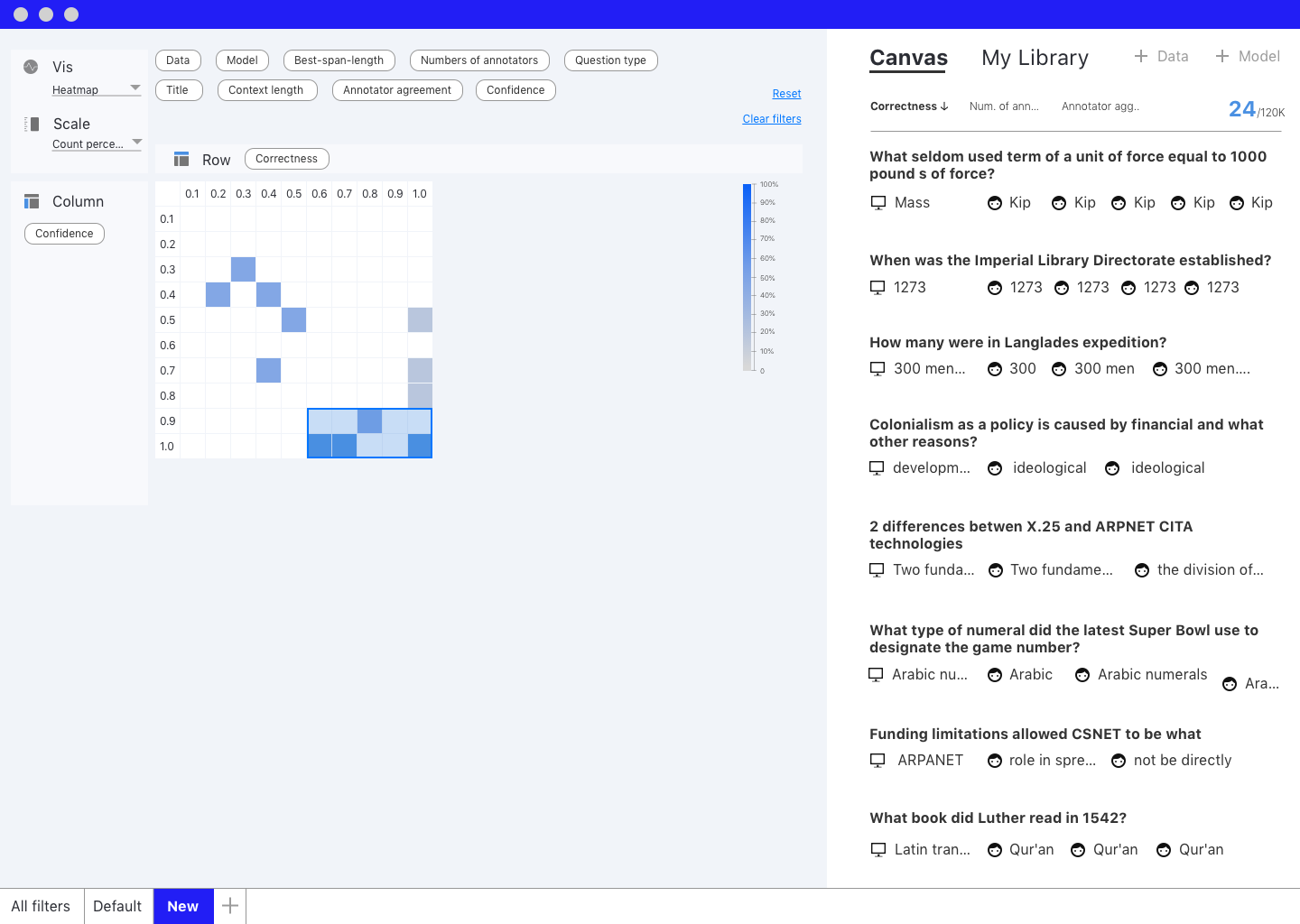}}
\caption{Round 2 Designs: (a)  The table was designed to show raw instance data and be sort-able along any column. Histograms on the left side allow the user to filter down to a subset of the data based on one or more attribute ranges. (b) The heatmap was designed to let the user see the relationship (joint distribution) between two variables. Brushing within the heatmap would reveal raw instance data for the selection.}
\label{fig:design-table}
\end{figure*}

\begin{figure*}[h]
\centering
\includegraphics[width=\textwidth]{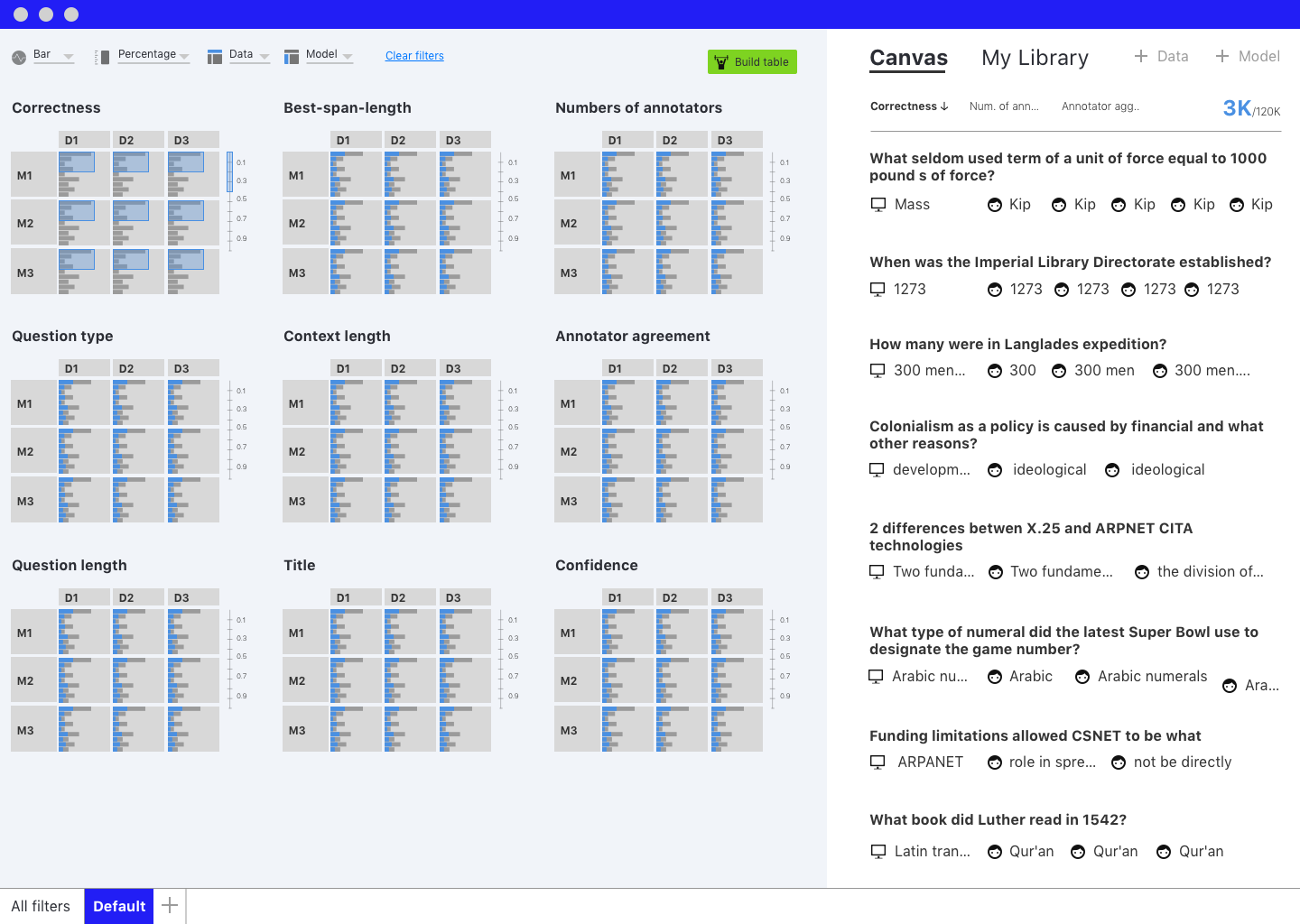}
\caption{Round 3 Designs: The confusion matrix visualization technique was extended by replacing cells with histograms of an additional variable. Selecting values ranges within this histogram would reveal raw instance data for the selection.}
\label{fig:design-crosscheck}
\end{figure*}

Figures~\ref{fig:design-table} and \ref{fig:design-crosscheck} supplement the design rounds discussed in Section~\ref{sec:design}. We produced these wireframes to elicit feedback from our users without implementing a full prototype.

\subsection{JavaScript and Python Packaging}

Determining an appropriate project structure was nontrivial. We packaged \cc as two separate but interdependent modules.
The visualization (JavaScript) portion of the code is a standalone \CodeFont{React}\footnote{\url{https://reactjs.org}} component packaged via NWB\footnote{\url{https://github.com/insin/nwb}}.
This retains the possibility of re-using the front-end of our tool in another web-based application with minimal re-engineering.

The second half of \cc is a python module that interfaces with the Jupyter notebook environment.
We started with \CodeFont{widget-cookiecutter}\footnote{\url{https://github.com/jupyter-widgets/widget-cookiecutter}} to follow best practices for Jupyter widget development, then augmented this code to  support \CodeFont{React} components.
This module also performs the data prepossessing (see Section~\ref{sec:jupyter-bottleneck} below) before transmitting the data from the python kernel to the web browser.
During the build process, the ``transpiled'' JavaScript code is copied into the python module, allowing the python module to be distributed and installed independently.

\end{document}